\pdfoutput=1
\documentclass{amsart}
\usepackage{amsmath,amsfonts,bm}
\usepackage{amssymb}
\usepackage{graphicx}
\usepackage{cite}

\begin{document}

\title[Oblique shock waves in wedge flow]{Analytic characterization of oblique shock waves in flows around wedges}

\author{Hady K. Joumaa}
\email{hkj@alum.mit.edu, hady.joumaa@gmail.com}

\date{Feb 13th, 2018}

\begin{abstract}
We consider the compressible flow around triangular wedges in which oblique shock waves are formed. We report on the novel analytic solution regarding the evaluation of the maximum wedge angle beyond which the shock wave detaches from the wedge to promote the formation of a bow shock. In addition, the limit line at which the flow past the oblique shock becomes sonic is determined whereby an analytic characterization for the corresponding shock angle is presented. 
\end{abstract}
\maketitle
\bibliographystyle{amsplain}

\section{Introduction}
\label{intro}
\begin{figure}
	\includegraphics[scale = 0.45]{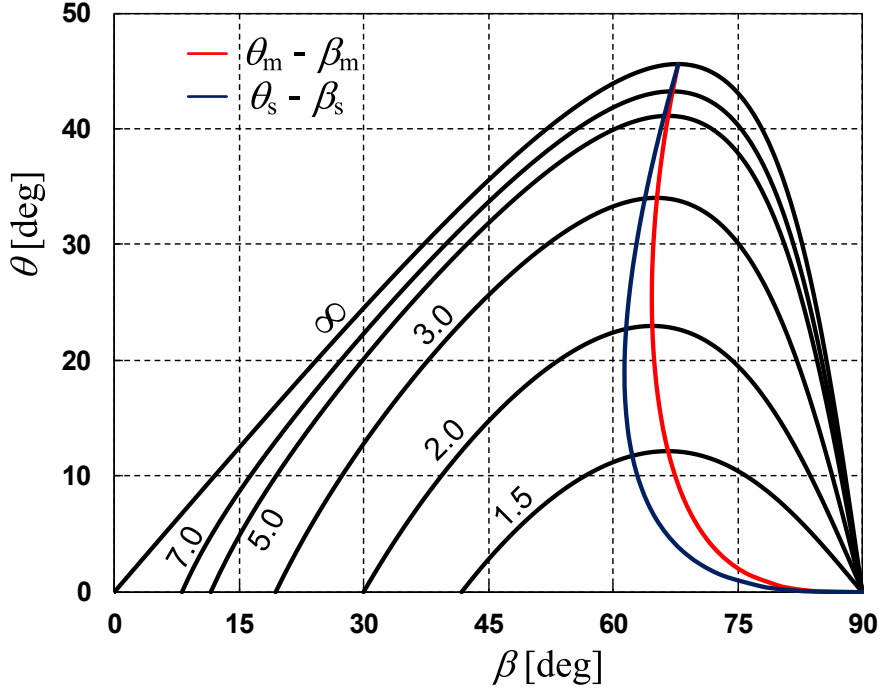}
	\caption{$\theta-\beta$ family curves for some special values of $M_1$. The sonic separatrix $\left( \theta_{\textnormal{s}} - \beta_{\textnormal{s}} \right)$  and the detachment separatrix $\left( \theta_{\textnormal{m}} - \beta_{\textnormal{m}} \right)$ are plotted. Note their unique intersection at $M_1 = 1$ and $M_1 \rightarrow \infty$. }
	\label{Th_Beta_curve_Family}	
\end{figure}

The supersonic flow past triangular wedge is discussed in many gas dynamics classics and is considered a renowned problem in the study of oblique shock wave in two-dimensional compressible flows \cite{Shapiro_BK, Thompson_BK, Anderson_BK}. Applying the fundamental conservation laws across the shock wave, the governing equation relating the free stream Mach number $M_1$, the wedge (deflection) angle $\theta$, and, the shock wave angle $\beta$, is formed. This equation is given as \cite{Anderson_BK, Emanuel_BK}
\begin{equation}
\label{Del_The_M_Eq}
	\tan \theta = \dfrac{2}{\tan \beta} \dfrac{M_1^2 \sin^2 \beta  -1}{M_1^2 \left( \gamma + \cos 2 \beta\right) + 2}
\end{equation}
$\gamma$ being the gas specific heat ratio. The resulting Mach number for the flow past the oblique shock wave, $M_2$, is calculated as follows \cite{Shapiro_BK}: 
\begin{equation}
\label{M_2_General}
	M_2^2 \sin^2 \left( \beta - \theta \right) = \dfrac{2 + \left(\gamma -1 \right) M_1^2  \sin^2 \beta}{2 \gamma M_1^2  \sin^2 \beta -\gamma + 1} 
\end{equation}
While Eq. \eqref{Del_The_M_Eq} appears to necessitate a numerical nonlinear solver if $\beta$ is to be solved in terms of $M_1$ and $\theta$, a clever substitution of the trigonometric functions in favour of $\tan \beta$ generates a cubic equation solvable in closed form by the method of radicals \cite{Emanuel_BK, Wolf_AR}. It is expressed as
\begin{multline}
\label{cubic_Th}
	\left[ 2 + \left( \gamma - 1\right) M_1^2 \right] \tan\theta \tan ^3\beta + 	2 \left(1 - M_1^2 \right) \tan ^2\beta \\
 + \left[ 2 + \left( \gamma + 1\right) M_1^2 \right] \tan\theta \tan \beta + 2 = 0
\end{multline}
Algebraically, Eq. \eqref{cubic_Th} admits three roots. The negative root is rejected and the remaining positive roots constitute the physical solution. The family $\theta - \beta$ curves for various $M_1$ are shown in Fig.~\ref{Th_Beta_curve_Family}. For any given $M_1 > 1$, the range of $\beta$ is bounded from below by the Mach angle, $\mu = \arcsin \left(\frac{1}{M_1}\right)$, and its upper bound is always $90 \deg$ corresponding to the case of normal shock wave. The solution curve of $\beta$ features a local maximum point at $\beta = \beta_{\textnormal{m}}$ with corresponding deflection angle denoted $\theta_{\textnormal{m}}$. For $\theta > \theta_{\textnormal{m}}$, the oblique shock detaches from the wedge and a bow shock is formed. The contour line joining the points $\left(\theta_{\textnormal{m}}, \beta_{\textnormal{m}}\right)$ constitutes a \emph{separatrix} splitting the region of the $\theta-\beta$ curves into a zone of strong shocks (situated to the right of separatrix) and a zone of weak shocks (situated to the left of separatrix) \cite{Anderson_BK}. In all previous studies, the $\left(\theta_{\textnormal{m}}, \beta_{\textnormal{m}}\right)$ points are determined by solving the nonlinear equation resulting from the straightforward method to determine local extrema points, i.e. set $\frac{d \theta}{d \beta} = 0$ and then solve for $\beta$ \cite{Wolf_AR}. In our work, we present a different approach to evaluate these maxima points. In fact, the first objective of this work is to formulate the analytic expressions of $\beta_{\textnormal{m}}\left( M_1, \gamma\right)$ and $\theta_{\textnormal{m}}\left( M_1, \gamma\right)$.

\begin{figure}
	\includegraphics[scale = 1.0]{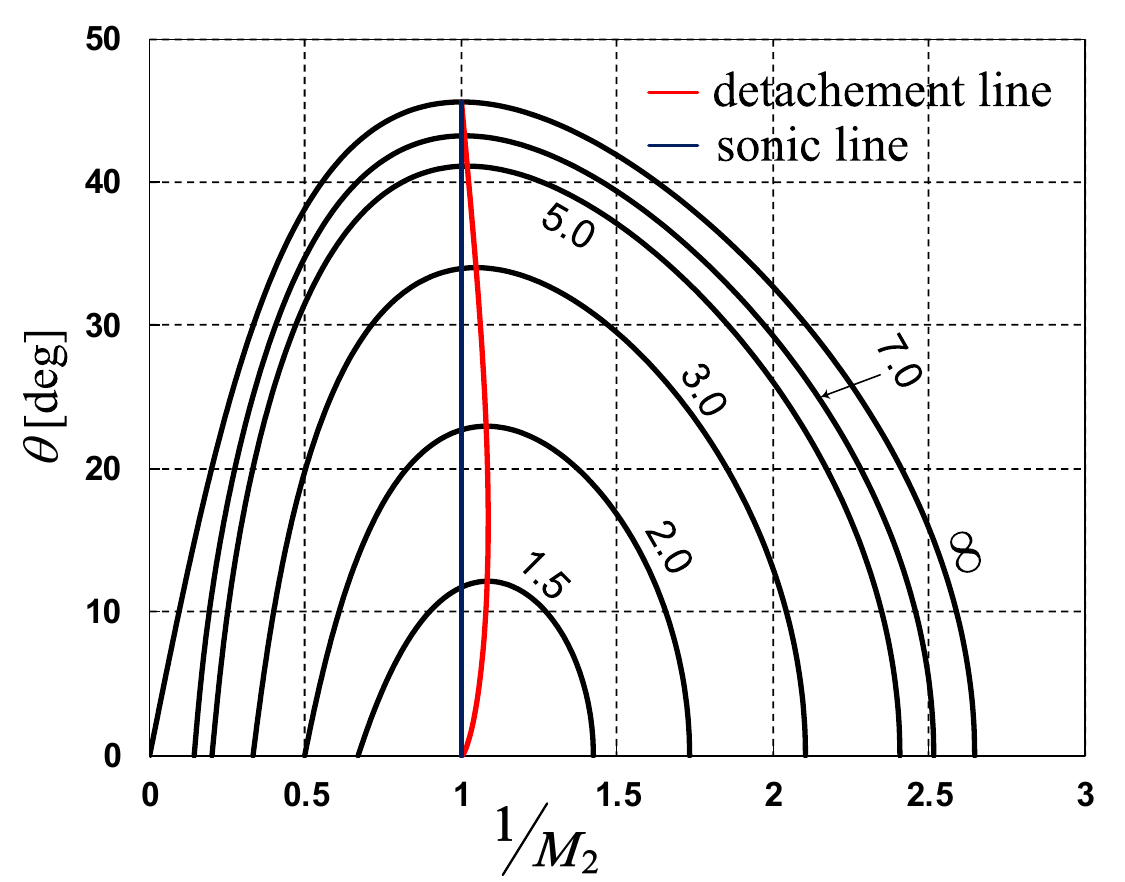}
	\caption{$\theta-M_2$ family curves for some special values of $M_1$. For convenience, $\frac{1}{M_2}$ is plotted. The sonic line corresponding to $M_2 = 1$ and the detachment line which joins all $\theta_{\textnormal{m}}$ points are also plotted.}
	\label{Th_M2_Family_curve}
\end{figure}

Another line of interest on the $\theta-\beta$ family curves is the sonic limit. On this line, the flow past the oblique shock is sonic, thus $M_2 = 1$. The sonic line intersects the $\theta-\beta$ curves at the particular points $\left( \theta_{\textnormal{s}}, \beta_{\textnormal{s}} \right)$. In similarity to the strong-weak shock segregation, the sonic limit separatrix splits the region of the $\theta-\beta$ family curves into a zone of supersonic flow to the left of separatrix and subsonic flow on its right side. The two separatrices intersect at two special points. The first, corresponds to $M_1 = 1$ whereby $ \beta_{\textnormal{m}} = \beta_{\textnormal{s}} = 90\deg$ and $\theta_{\textnormal{m}} = \theta_{\textnormal{s}} = 0\deg$; in fact for the trivial case $M_1 = 1$, the entire $\theta-\beta$ curve collapses to the one point $\left(0\deg, 90\deg \right)$. The second point of intersection corresponds to $M_1 \rightarrow \infty$, in such a case, we obtain $ \beta_{\textnormal{m}} = \beta_{\textnormal{s}} \rightarrow \arctan \sqrt{\frac{\gamma +1}{\gamma-1}}$ and $ \theta_{\textnormal{m}} = \theta_{\textnormal{s}} \rightarrow \arctan \sqrt{\frac{1}{\gamma^2 -1}}$, these results are derived in the upcoming sections. We also prove that for all physically meaningful problems, $\beta_{\textnormal{m}} \geq \beta_{\textnormal{s}}$ and $\theta_{\textnormal{m}} \geq \theta_{\textnormal{s}}$; equality occurs only at the two limiting points just introduced. From this inequality, we deduce that strong shocks always result in a subsonic flow, while flows past weak shocks can either be subsonic or supersonic. This conclusion is graphically illustrated in Fig.~\ref{Th_M2_Family_curve} whereby the variation of $\theta$ with respect to $\frac{1}{M_2}$ is plotted. The detachment separatrix is situated in the zone $\frac{1}{M_2} > 1$, corresponding to subsonic downstream flow. The zone between the two separatrices corresponds to weak shocks albeit a subsonic downstream flow.

\section{Closed form evaluation of $\beta_{\textnormal{m}}$ and $\theta_{\textnormal{m}}$}
\label{sec:1}
\begin{figure}
	    \includegraphics[scale = 0.40]{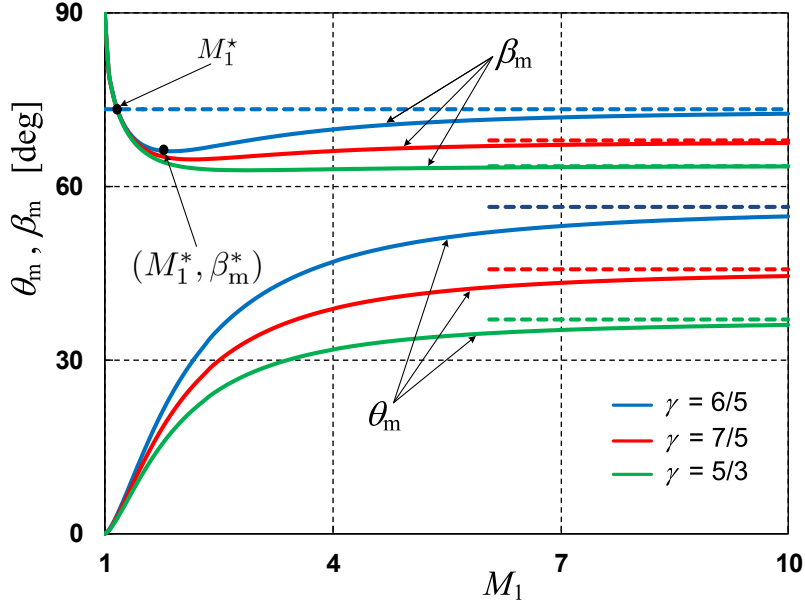}
		\caption{The variation of $\beta_{\textnormal{m}}$ and $\theta_{\textnormal{m}}$ with respect to $M_1$. The asymptotes for large $M_1$ are also shown. The local minima points on the $\beta_{\textnormal{m}}$ curve $\left( M_1^{\ast}, \beta_{\textnormal{m}}^{\ast} \right)$ are indicated.}
\label{Th_Be_Max_curve}	
\end{figure}	
A general cubic equation of the form $ax^3 + bx^2 + cx + d = 0$ admits a double root $x_{\textnormal{d}}$ if its discriminant vanishes. The double root also nullifies the first derivative of the polynomial equation. Therefore, 
\begin{subequations}
\begin{equation}
\label{cubic_condition}
27 a^2 d^2 - b^2c^2 + 4 ac^3 - 18 abcd + 4b^3d = 0
\end{equation}
\begin{equation}
\label{double_root_form}
3ax_{\textnormal{d}}^2 + 2bx_{\textnormal{d}} + c = 0
\end{equation}
\end{subequations}
Applying the formula of Eq. \eqref{cubic_condition} to the coefficients of Eq. \eqref{cubic_Th}, the result is a mathematical characterization of the local maximum point $\left(\theta_{\textnormal{m}},  \beta_{\textnormal{m}} \right)$ and consequently, an algebraic equation with $\tan\theta_{\textnormal{m}}$ as unknown emerges. Thus
\begin{multline}
\label{theta_max_Eq}
\left[ 2 + \left(\gamma - 1\right)M_1^2 \right] \left[ 2 + \left(\gamma + 1\right)M_1^2 \right]^3 \tan^4 \theta_{\textnormal{m}} \\
 + \Bigl \lbrace 27\left[ 2 + \left(\gamma - 1\right)M_1^2 \right]^2 -  \left(1 - M_1^2\right)^2\left[ 2 + \left(\gamma + 1\right)M_1^2 \right]^2  \\
 - 18\left(1 - M_1^2\right) \left[ 2 + \left(\gamma - 1\right)M_1^2 \right] \left[ 2 + \left(\gamma + 1\right)M_1^2 \right]  \Bigr \rbrace  \tan^2 \theta_{\textnormal{m}} \\
  + 16\left( 1 - M_1^2\right)^3 = 0 
\end{multline}
Eq. \eqref{theta_max_Eq} is of bi-quadratic type with only one root being physically meaningful (the product of all four roots is negative). In conjunction with the admissible solution of Eq. \eqref{theta_max_Eq}, a closed form expression for $\cos 2 \theta_{\textnormal{m}}$ is provided in \cite{Wolf_AR}. In the limiting case $M_1 \rightarrow \infty$, the physical solution approaches the asymptotic value $\sqrt{\frac{1}{\gamma^2 - 1}}$. The solution $ \theta_{\textnormal{m}} \left( M_1\right)$ for various values of $\gamma$ is plotted in Fig.~\ref{Th_Be_Max_curve}; all plots exhibit identical behaviour regarding the consistent increase of $ \theta_{\textnormal{m}}$ toward its asymptotic value. 

Considering Eq. \eqref{double_root_form} and Eq. \eqref{cubic_Th} and eliminating $\tan \theta_{\textnormal{m}}$, we obtain the governing equation for $\tan \beta_{\textnormal{m}}$ which takes the following form 
\begin{multline}
\label{Th_max_Eq}
\left(1 - M_1^2 \right) \dfrac{ 2 + \left( \gamma - 1\right)M_1^2 }{2 + \left( \gamma + 1\right)M_1^2} \tan^4 \beta_{\textnormal{m}} \\
+\dfrac{ \left( \gamma + 1\right) M_1^4 +2 \left( \gamma - 1\right) M_1^2 + 4 }{2 + \left( \gamma + 1\right)M_1^2} \tan^2 \beta_{\textnormal{m}} + 1 = 0
\end{multline}
This is also a bi-quadratic equation in $\tan\beta_{\textnormal{m}}$, admitting one admissible solution expressed as
\begin{multline}
\label{Th_max_Sol}
\dfrac{\tan^2 \beta_{\textnormal{m}}}{M_1^2} = \dfrac{\sqrt{\left( \gamma  +1\right) \left[ \left(M_1^2 - 4 \right)^2 + M_1^2 \gamma  \left(M_1^2 +8 \right)\right] }}{2\left(M_1^2 - 1 \right) \left[2 + \left( \gamma - 1\right)M_1^2 \right]} \\
+ \dfrac{\left( \gamma + 1\right) M_1^2 +2 \left( \gamma - 1\right) + \dfrac{4}{M_1^2} }{2\left(M_1^2 - 1 \right) \left[2 + \left( \gamma - 1\right)M_1^2 \right]}
\end{multline}
From Eq. \eqref{Th_max_Sol}, it is easy to verify that $\lim\limits_{M_1\rightarrow \infty}\tan \beta_{\textnormal{m}} = \sqrt{\frac{\gamma  +1}{\gamma - 1}} $. Interestingly, $\beta_{\textnormal{m}}$ crosses this limit at a finite value of $M_1$ denoted $M_1^{\star}$. From Fig.~\ref{Th_Be_Max_curve} where $\beta_{\textnormal{m}}$ is plotted for various $\gamma$, we notice that $\beta_{\textnormal{m}}$ decreases from $90\deg$ toward a local minimum point ($M_1^{\ast}, \beta_{\textnormal{m}}^{\ast}$) and then approaches its asymptote from below. The intersection of $\beta_{\textnormal{m}}$ with its asymptote corresponds to $M_1^{\star}$. In the following, we evaluate $M_1^{\star}$, $M_1^{\ast}$, and $\beta_{\textnormal{m}}^{\ast}$.

To evaluate $M_1^{\star}$, we consider Eq. \eqref{Th_max_Eq} and substitute for $\tan^2 \beta_{\textnormal{m}}^{\star} = \frac{\gamma  +1}{\gamma - 1}$. The result is a bi-quadratic equation in $M_1$ having a single meaningful root $M_1^{\star}$ given by       
\begin{equation}
\label{Th_max_inter}
M_1^{\star} = \sqrt{\dfrac{2 \gamma}{\left( \gamma  +1\right) \left(2 - \gamma \right)}}
\end{equation}
The evaluation of $M_1^{\ast}$, and $\beta_{\textnormal{m}}^{\ast}$ requires the rewriting of Eq. \eqref{Th_max_Eq} as to make $M_1$ the unknown and $\tan \beta_{\textnormal{m}}$ the parameter. For simplicity, we substitute $\tan^2 \beta_{\textnormal{m}}$ by $t$ and $M_1^2$ by $m$. The treated equation becomes
\begin{multline}
\label{Param_Eq}
\left[ -t^2\left(\gamma - 1 \right) + t\left(\gamma + 1 \right)\right] m^2 \\
+\left[ \left(\gamma - 3 \right)t^2 + 2\left(\gamma - 1\right)t + \gamma + 1 \right] m + 2\left(t +1\right)^2 = 0
\end{multline}
The local minimum point, ($M_1^{\ast}, \beta_{\textnormal{m}}^{\ast}$) corresponds to Eq. \eqref{Param_Eq} having a double root in $m$. Indeed, for $\tan^2 \beta_{\textnormal{m}}^{\ast} < t < \frac{\gamma  +1}{\gamma - 1}$, there exists two distinct roots for $m$. For Eq. \eqref{Param_Eq} to have a double root, its discriminant must vanish. The discriminant is $\left(t  +1 \right)^2 \left[ \left(\gamma  +1 \right)t^2 + 2 \left( \gamma - 7\right)t + \gamma  +1 \right]$ and the three values of $t$ that nullify it are $t_1 = -1$, $t_2 = \frac{\left(\sqrt{3 - \gamma} -2 \right)^2}{\gamma+1}$, and $t_3 = \frac{\left(\sqrt{3 - \gamma} + 2 \right)^2}{\gamma+1}$. $t_1$ being negative, does not correspond to a physical solution. $t_2$ and $t_3$ are both positive, nevertheless, $t_2$ generates a double root $m_{\textnormal{d}} < 1$ and this is unacceptable since $M_1 > 1$. Thus, $t_3$ will be the only acceptable solution corresponding to     
\begin{subequations}
\begin{align}
\begin{split}
M_1^{\ast} = & \, 2 \sqrt{\dfrac{\gamma -1+ \sqrt{3 - \gamma}}{\left( \gamma+1\right)\left(2- \gamma\right)}}
\end{split} \\
\begin{split}
\tan \beta_{\textnormal{m}}^{\ast} = & \, \dfrac{\sqrt{3 - \gamma} + 2}{\sqrt{\gamma+1}} 
\end{split}
\end{align}
\end{subequations}

\section{Closed form evaluation of $\beta_{\textnormal{s}}$}
\begin{figure}
	\includegraphics[scale = 0.40]{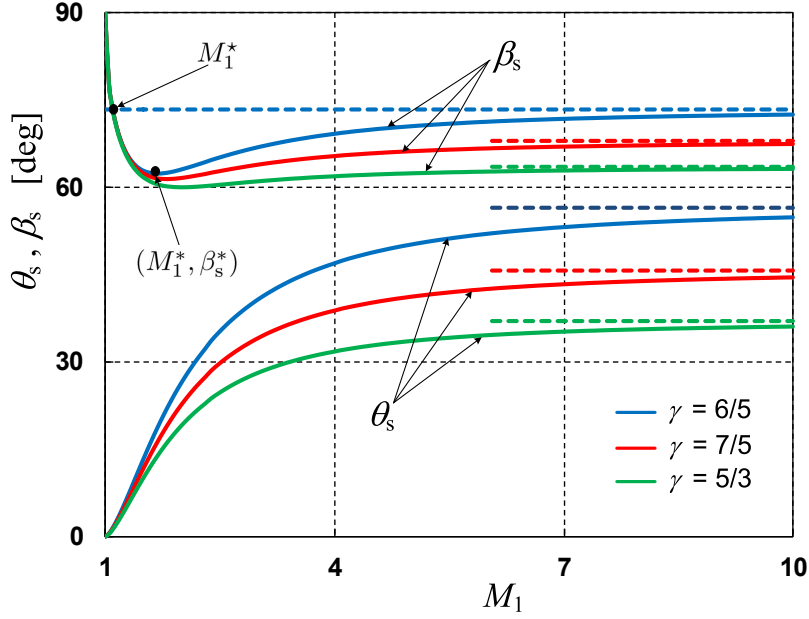}
	\caption{The variation of $\beta_{\textnormal{s}}$ and $\theta_{\textnormal{s}}$ with respect to $M_1$. The asymptotes for large $M_1$ are also shown. The local minima points on the $\beta_{\textnormal{s}}$ curve $\left( M_1^{\ast}, \beta_{\textnormal{s}}^{\ast} \right)$ are indicated.}
	\label{Th_Be_Sonic_curve}	
\end{figure}	
Setting $M_2 = 1$ in Eq. \eqref{M_2_General}, a relation between $\beta_{\textnormal{s}}$ and $\theta_{\textnormal{s}}$ emerges 
\begin{equation}
\label{Th_Del_Sonic}
	\sin^2 \left( \beta_{\textnormal{s}} - \theta_{\textnormal{s}} \right) = \dfrac{2 + \left(\gamma -1 \right) M_1^2  \sin^2 \beta_{\textnormal{s}}}{2 \gamma M_1^2  \sin^2 \beta_{\textnormal{s}} -\gamma + 1} 
\end{equation}
The relation of Eq. \eqref{Th_Del_Sonic}, along with that of Eq. \eqref{Del_The_M_Eq} constitute the incomplete system of nonlinear equations whose unknowns are $\beta_{\textnormal{s}}$, $\theta_{\textnormal{s}}$, and $M_1$. After extensive algebraic work aiming to eliminate $\theta_{\textnormal{s}}$, a bi-quadratic equation with unknown $\tan\beta_{\textnormal{s}}$ and parameter $M_1$ is obtained. Hence,  
\begin{multline}
\label{Th_Sonic_Biquad}
\left(1 - M_1^2 \right) \left[2 + \left( \gamma - 1\right)M_1^2 \right] \tan^4 \beta_{\textnormal{s}} \\
 +\left[ M_1^4 \left( \gamma + 1\right) + M_1^2 \left( \gamma - 3 \right)  +4 \right] \tan^2 \beta_{\textnormal{s}}  + 2 = 0 
\end{multline}
and the physically meaningful root of Eq. \eqref{Th_Sonic_Biquad} is
\begin{equation}
\label{Th_Sonic_Exp}
\dfrac{\tan^2 \beta_{\textnormal{s}}}{M_1^2} = \dfrac{ \sqrt{\left[ M_1^2\left(\gamma+1 \right) + \gamma - 3 \right]^2 + 16\gamma} + M_1^2 \left( \gamma + 1\right) + \gamma - 3 +\frac{4}{M_1^2}}{2\left(M_1^2  -1 \right) \left[2 + \left(\gamma - 1\right)M_1^2 \right]}
\end{equation}
Having solved analytically for $\beta_{\textnormal{s}} \left( M_1, \gamma\right)$, the expression for $\theta_{\textnormal{s}} \left( M_1, \gamma\right)$ is derived by engaging Eq. \eqref{Del_The_M_Eq}. The plots of $\theta_{\textnormal{s}}\left( M_1, \gamma\right)$ and $\beta_{\textnormal{s}} \left( M_1, \gamma \right)$ for three distinct values of $\gamma$ are shown in Fig. \ref{Th_Be_Sonic_curve}. In the limiting case $M_1 \rightarrow \infty$, the root in Eq. \eqref{Th_Sonic_Exp}, i.e. $\tan\beta_{\textnormal{s}}$, approaches $\sqrt{\frac{\gamma+1}{\gamma -1}}$ while the asymptotic limit for $\tan \theta_{\textnormal{s}}$ is $\sqrt{\frac{1}{\gamma^2 - 1}}$. This latter limit is obtained by substituting for the appropriate values of $M_1$ and $\beta_{\textnormal{s}}$ in Eq. \eqref{Del_The_M_Eq}.

In the plots of Fig. \ref{Th_Be_Sonic_curve}, we highlight on two interesting points: the first corresponding to the intersection of the curve with its asymptote, and the second being the local minimum. The first point is evaluated by substituting for the asymptotic value in Eq. \eqref{Th_Sonic_Biquad} to obtain $M_1^{\star} = 2 \sqrt{\frac{\gamma}{\left(\gamma  +1 \right)\left(3 - \gamma \right)}}$. The local minimum is determined by first rewriting Eq. \eqref{Th_Sonic_Biquad} in which the unknown becomes $m = M_1^2$ and the parameter is $t = \tan^2\beta_{\textnormal{s}}$. Hence, the following quadratic equation is produced
\begin{equation}
\label{M_Sonic_Biquad}
\left[ \left( \gamma - 1\right) t^2 - \left( \gamma + 1\right)t \right] m^2 
+ \left(3 -  \gamma \right) \left(t^2 + t \right) m - 2 \left(t + 1\right)^2 = 0
\end{equation}      
The local minimum $\left( M_1^{\ast}, \beta_{\textnormal{s}}^{\ast}\right)$ coincides with the point at which Eq. \eqref{M_Sonic_Biquad} admits a double root. The caveat for double root along with its value correspond to the following results
\begin{subequations}
\begin{align}
\begin{split}
\beta_{\textnormal{s}}^{\ast} = & \,\arctan \sqrt{\dfrac{8}{\gamma+1}}
\end{split} \\
\begin{split}
M_1^{\ast} = & \, \sqrt{\dfrac{\gamma + 9}{2 \left(3 - \gamma\right)}}
\end{split} 
\end{align}
\end{subequations}

\section{Proof of $\beta_{\textnormal{m}} > \beta_{\textnormal{s}}$}
The aim of the following work is to prove that $\beta_{\textnormal{m}} > \beta_{\textnormal{s}}$ for all admissible $\gamma$ and $M_1$. Considering the two equations that solve for $t_\textnormal{m} = \tan^2\beta_{\textnormal{m}} $ and $t_\textnormal{s} = \tan^2\beta_{\textnormal{s}}$, mainly Eqs. \eqref{Th_max_Eq} and \eqref{Th_Sonic_Biquad}, we realise that they are both of the form $At^2 + Bt + C = 0$ sharing a common $A$ term but with different $B$ and $C$. The expressions for these coefficients along with the roots of interest $t_{\textnormal{m}}$ and $t_{\textnormal{s}}$, are given in Eq. \eqref{Coeff_Exp}.
\begin{subequations}
\label{Coeff_Exp}
\begin{align}
\begin{split}
A = & \left(1 - M_1^2\right) \left[2 + \left(\gamma-1 \right)M_1^2 \right] < 0
\end{split} \\
\begin{split}
B_\textnormal{m} - B_\textnormal{s} =& \,C_\textnormal{m} - C_\textnormal{s} = 
M_1^2\left(\gamma+1\right) > 0
\end{split}\\
\begin{split}
B_\textnormal{m} + B_\textnormal{s} = & \,2\left(\gamma+1\right)M_1^4 + \left(3\gamma - 5\right)M_1^2 + 8
\end{split}\\
\begin{split}
t_\textnormal{m} =& \dfrac{-B_\textnormal{m} - \sqrt{B_\textnormal{m}^2 - 4AC_\textnormal{m}}}{2A}
\end{split}\\
\begin{split}
t_\textnormal{s} =& \dfrac{-B_\textnormal{s} - \sqrt{B_\textnormal{s}^2 - 4AC_\textnormal{s}}}{2A}
\end{split}
\end{align}
\end{subequations} 
To prove that $\beta_{\textnormal{m}} > \beta_{\textnormal{s}}$ is equivalent to prove $t_\textnormal{m} > t_\textnormal{s}$. From the expressions of $t_\textnormal{m}$ and $t_\textnormal{s}$, we have
\begin{align*}
 & \, \dfrac{-B_\textnormal{m} - \sqrt{B_\textnormal{m}^2 - 4AC_\textnormal{m}}}{2A} > \dfrac{-B_\textnormal{s} - \sqrt{B_\textnormal{s}^2 - 4AC_\textnormal{s}}}{2A} \\ 
 \Leftrightarrow & \, -B_\textnormal{m} - \sqrt{B_\textnormal{m}^2 - 4AC_\textnormal{m}} < -B_\textnormal{s} - \sqrt{B_\textnormal{s}^2 - 4AC_\textnormal{s}} \\ 
 \Leftrightarrow & \, \sqrt{B_\textnormal{s}^2 - 4AC_\textnormal{s}} - \sqrt{B_\textnormal{m}^2 - 4AC_\textnormal{m}} < B_\textnormal{m} - B_\textnormal{s}  \\  
\Leftrightarrow & \dfrac{B_\textnormal{s}^2 - 4AC_\textnormal{s} - B_\textnormal{m}^2 + 4AC_\textnormal{m}}{\sqrt{B_\textnormal{s}^2 - 4AC_\textnormal{s}} + \sqrt{B_\textnormal{m}^2 - 4AC_\textnormal{m}}} < B_\textnormal{m} -B_\textnormal{s} \\
\Leftrightarrow & \, \dfrac{4A\left( C_\textnormal{m} - C_\textnormal{s}\right) - \left( B_\textnormal{m} - B_\textnormal{s}\right) \left(B_\textnormal{m} + B_\textnormal{s} \right)}{B_\textnormal{m} -B_\textnormal{s}} < \sqrt{B_\textnormal{s}^2 - 4AC_\textnormal{s}} + \sqrt{B_\textnormal{m}^2 - 4AC_\textnormal{m}} \\ 
  \Leftrightarrow & \, 4A  -  \left(B_\textnormal{m} + B_\textnormal{s} \right) <  \sqrt{B_\textnormal{s}^2 - 4AC_\textnormal{m}} + \sqrt{B_\textnormal{m}^2 - 4AC_\textnormal{m}} \\
 \Leftrightarrow & \, M_1^2 \left[\left(\gamma - 7 \right) + 2\left(1 - 3 \gamma \right) M_1^2\right] <  \sqrt{B_\textnormal{s}^2 - 4AC_\textnormal{m}} + \sqrt{B_\textnormal{m}^2 - 4AC_\textnormal{m}} 
\end{align*}
the last inequality is always satisfied since the right hand side is positive and the left hand side is negative. Thus the initial assumption $t_\textnormal{m} > t_\textnormal{s}$ is true and consequently $\beta_{\textnormal{m}} > \beta_{\textnormal{s}}$ for all $M_1>1$. A direct deduction to this result is $\theta_{\textnormal{m}} > \theta_{\textnormal{s}}$ which implies the feasibility of having weak shock with subsonic downstream flow.  
\section{Conclusion}
In this paper, the oblique shock problem is revisited and the closed form expressions for angles of shocks corresponding to sonic limit and detachment limit are formulated. These angles are further characterised whereby their asymptotic limit and local minimum point are evaluated analytically. The followed mathematical procedure is based on seeking the multiple roots of third and fourth order algebraic parametric equations. This approach lead to the composition of a concise proof for the restriction relation $\beta_{\textnormal{m}} > \beta_{\textnormal{s}}$.


\begin{thebibliography}{99}
 
\bibitem{Shapiro_BK}
A. Shapiro, \emph{The Dynamics and Thermodynamics of Compressible Fluid Flow}, The Ronald Press Company, V.I, 1953. 

\bibitem{Thompson_BK}
P. A. Thompson, \emph{Compressible Fluid Dynamics}, McGraw Hill, 1972. 

\bibitem{Anderson_BK}
J. D. Anderson, \emph{Modern Compressible Flow}, McGraw Hill, 3rd edition, 2003. 

\bibitem{Emanuel_BK}
G. Emanuel, \emph{Analytical Fluid Dynamics}, CRC Press, 2016. 

\bibitem{Wolf_AR} 
T. Wolf, Comment on "approximate formula of weak oblique shock wave angle", \emph{AIAA Journal}, \textbf{31}(7) (1993), pp. 1363-1363.
\end{thebibliography}
\end{document}